\documentclass[12pt]{article}

\date {3 April 2007}

\begin{document}

\title{Astroelectrons as Non-Dual Field Solutions}  

\author {I.E. Bulyzhenkov-Widicker}

\maketitle

\begin{abstract}
The continuous charge density of the $r^{-4}$  radial astroelectron is found analytically from the Maxwell-Lorentz equations. The electric charge is not a basic concept, but is the nonlinear field distribution associated with the elementary self-energy density. The Poisson equation for the unified field carrier in non-dual electrodynamics locally relates the post-Coulomb potential with the  energy-charge density. The local Maxwell equations maintain the global overlap of nonlocal electric energy-charges in the undivided Universe.  

 

\end{abstract}

\bigskip

For a long time the point microscopic electron for the Maxwell-Lorentz equations \cite {Lor}, $\nabla {\bf d}(x) = 4\pi \rho(x), \  c \nabla\times{\bf b}(x)  = 4\pi \rho (x){\bf v} + \partial_t  {\bf d}(x), \ \nabla {\bf b}(x)$ $ = 0, $ $
 \ c\nabla \times {\bf d} (x)  = - \partial_t {\bf b}(x)$, has been considered as `an attempt which we have called intellectually unsatisfying' \cite{Bro}. Einstein called directly for non-dual physics of classical fields and their extended sources: `A coherent field theory requires that all elements be continuous... And from this requirement arises the fact that the material particle has no place as a basic concept in a field theory. Thus, even apart from the fact that it does not include gravitation, Maxwell's theory cannot be considered as a complete theory' (translation \cite{Ton}). 

Fortunately for Classical Electrodynamics (CE), the continuous astroelectron does belong to Max\-well's equations, which may ri\-gorously suggest a rotationally invariant energy-type solution for the electrical charge density,  $ \rho (r) =  r_o{\bf d}^2(r)/4\pi  q $ $\equiv$ $ 2r_o w(r)/q$, of a radial source $q=\int_o^\infty 4\pi r^2 \rho (r)dr$ in its frame of references (${\bf v} = {\bf b} =0$), 
\begin {eqnarray}
\cases { 
\rho (r)  = qr_o /4\pi  r^2(r + r_o)^2 =  \nabla {\bf d}(r) / 4\pi \cr\cr
{\bf d}(r) = {q{\hat {\bf r}}/  r(r+r_o)} = - \nabla \varphi (r)  \cr\cr
\varphi (r) = (q/r_o) ln [(r+r_o)/r]\cr\cr
W = \int dV {\bf d}^2/8\pi  =  {q}\int dV \rho/2r_o = q^2/2r_o  \equiv \int\!dV \rho \varphi/2, \cr  
}\end  {eqnarray}
where q = $-e_o$ is the electron charge with a half of its value within the phenomenological radius $r_o$. It is in full agreement with the Gauss integral theorem and the vector CE identities  that the electrostatic field self-energy, $\int\!dV{\bf d}^2/8\pi$, is equal to the electrostatic charge self-energy, $\int\!dV \rho \varphi/2$, with the post-Coulomb potential $\varphi(r) \propto ln (1 + r_or^{-1})$. What is really new in (1) is that the analytical CE solution maintains the finite  self-energy $W$, justified \cite{Sch} for the charged source by Quantum Electrodynamics(QE).  Moreover, CE reveals in the exact solution (1) that the continuous source of Maxwell fields is locally proportional to their energy density distributed over the entire Universe.

One might assume for a moment that there is no much sense to develop extended classical sources, including the $r^{-4}$ radial astroelectron (1), because any theory without quantization would never compete with the tremendous QE achievements for the Dirac electron. But why Schwinger, the founder of Source Theory and Quantum Electrodynamics, studied classical radial sources and dedicated to Dirac the CE approach \cite{Sch1} to the distributed electron?  In our view, both Dirac and Schwinger anticipated that CE can rigorously employ astrospace distributions for electric charges and clarify their relativistic energy nature. Contrary to quantum probabilities, classical fields may change the empty-space paradigm of modern physics and address many conceptual source problems in gravitation and astrophysics. Classical and quantum approaches to matter should tend to support and replicate each other because they both have to describe the  same physical reality. Therefore a complete non-dual theory of classical fields should replicate in one or another way the proved nonlocality of quantum matter.

The weak-field electric potential in (1), $\varphi (r)
\approx (q/r) - (qr_o/2r^2)$  for $r_o \leq r$, may take the exact Coulomb limit
for the `point particle in empty space'  model, where $r_o \equiv 0$. And the Coulomb potential $q/r$ is the solution of the Laplace equation for the empty space, when $\nabla^2\varphi = 0$, rather than of the Poisson equation for the non-empty space with continuous distribution of the elementary charge, when $\nabla^2\varphi \neq 0$. The "point particle in empty space" modeling of physical reality has transformed Maxwell's electrodynamics into the incomplete theory with the energy divergence problem and the ill-defined (not ill-derived) source. However non-empty space solutions for continuous elementary sources may rigorously solve this conceptual problem for classical fields, for example through the non-linear Poisson equation for the radial electron (1),
\begin {equation}
 -{1\over 4\pi}\nabla^2 \varphi (r) = {r_o\over 4\pi q} [\nabla \varphi (r)]^2 \equiv \rho(r).
 \end {equation}
 Here one may relate traditionally the linear field term to the classical field of the continuous carrier, while the non-linear (quadratic) field term to its particle-source matter or the charge-energy density $\rho$. In fact, the continuous astroelectron (1) is a carrier of linear and non-linear self-field densities, which are locally bound  at the same space point due to the same argument at the right- and left-hand sides of (2). Therefore both these continuous densities of one elementary carrier might be described in covariant presentations by the same four-potential $a_i(x)$ with the traditional decomposition of electricity into its field and current parts. One should not anymore assign field and source functions of the electron to different mathematical arguments (or to formal delta operators with
 different space coordinates) in opposite sides of the analytical Poisson  equation.

Our next goal can be to design a covariant option of the nonli\-near electrostatic equation (2) for the self-potential of the elementary astrocarrier of electricity. We take  into consideration that Maxwell's equations and Lorentz's field transformations cannot admit nonlinear functions (with the self-potential $a_i = \{\varphi, {\bf a}\}$) in the classical field $f_{ik} (x) = \nabla_i a_k(x) - \nabla_k a_i(x)$.  Therefore we will stay with the classical definition of the electromagnetic field tensor $f_{ik}$  and with the Maxwell linear relations, $\nabla_l f_{ik}(x) + \nabla_i f_{kl}(x) + \nabla_k f_{li}(x) = 0$, between the self-field components $f_{ik}$ of the  $r^{-4}$ radial source (1).

  The right hand side of the non-dual field equation (2) is not a linear function with respect to the local self-potential $a_i(x)$ or the self-field intensities $\nabla_k a_i$. Here one may gather all non-linear field terms for the conventional introduction of the elementary current density, $j^i(x)$, that makes sense for further descriptions of the multi-carrier world ensemble of continuous particles.
Recall that  the Lorentz  microscopic electron was used to model macroscopic current densities in a medium \cite{Lor}. One may also expect a transition from the nonlinear field equation for one elementary astrocarrier to linear equations for the classical combination of net field and charge densities.  Indeed, a linear superposition of overlapping densities (2) is possible only for all field intensities, $\sum^\infty_n\nabla_kf_n^{ki}(x)$ =
$\nabla_k\sum^\infty_nf_n^{ki}(x)$ $\equiv \nabla_kF^{ki}(x)$, 
 while nonlinear intensities (elementary energy-fields) can not comply to the superposition principle. Therefore an extra phenomenological  density, a current of electric energy-charges $J^i(x)\equiv  \sum^\infty_n j_n^i(x)$), should be defined for the local overlap of elementary non-linear functions at joint space-time points $x$. Then one may manifest  
  a local electrodynamic equation, $\nabla_kF^{ki}(x) = 4\pi c^{-1} J^i(x) $, for the world ensemble of overlapping electric charges-energies and their fields.
 
The conceptual linearity of Maxwell's  electrodynamics with respect to both field and current densities reflects the  superposition criterion for  distributed energy or charge densities, rather than reflects a unique, non-field nature of the elementary charge.  The energy divergence-free astrocarrier of electricity can be analytically introduced as a pure field or non-dual object with local relations between non-linear functions of field intensities $f_{ik}$ and their first derivatives. The phenomenological current or charge density is not a basic CE concept, but is the local overlap of all nonlinear field distributions or nonlocal elementary energies. In closing, the exact analytical  solution (1) for Maxwell's equations requests the $r^{-4}$ radial astroparticle  and the non-empty space paradigm for gravitation.

\end {document}